\documentclass[notitlepage,showkeys,onecolumn,11pt,times,superscriptaddress,a4paper,pre,aps]{revtex4-2}
\usepackage{stix}
\usepackage[T1]{fontenc}
\usepackage[english]{babel}
\usepackage{graphicx}
\usepackage{bm}
\usepackage[colorlinks=true,
            linkcolor=blue,
            urlcolor=blue,
            citecolor=blue]{hyperref}
\usepackage[mathlines]{lineno}
\usepackage{amsmath}
\usepackage{physics}
\usepackage{booktabs}
\usepackage{geometry}
\begin{document}


\title{Stochastic-layer width as an order parameter for the integrability-to-chaos transition in the bouncer model}

\author{Daniel Borin}
\affiliation{S\~ao Paulo State University (UNESP), Institute of Geosciences and Exact Sciences, 13506-900, Rio Claro, SP, Brazil}

\author{Arturo C. Marti }
\affiliation{Universidad de la República, Facultad de Ciencias, Igua 4225, Montevideo, Uruguay}

\author{Edson D. Leonel}
\affiliation{S\~ao Paulo State University (UNESP), Institute of Geosciences and Exact Sciences, 13506-900, Rio Claro, SP, Brazil}


\begin{abstract}
The transition from integrable to chaotic dynamics in the conservative bouncer model is investigated within the phenomenological framework of continuous phase transitions. In the integrable limit, $\epsilon = 0$, the action variable is conserved and the phase space is completely foliated by horizontal invariant curves. Introducing a weak periodic perturbation, $\epsilon > 0$, destroys resonant tori and generates localized stochastic layers around the fixed points of the mapping, even at parameter values where global phase-space portraits appear predominantly regular. By dividing the angular domain into discrete bins and extracting the extreme action values visited by a single chaotic trajectory, the upper and lower boundary curves of the stochastic layers surrounding the period-one fixed points $(\pi, 0)$ and $(\pi, \pi)$ are reconstructed numerically. For the fixed point $(\pi, 0)$, the stochastic layer develops in the region $V \geq 0$ and is bounded by two curves, whereas around $(\pi, \pi)$ the layer extends across both sides of the line $V = \pi$ and requires a four-boundary characterization. In both cases, the mean width of the stochastic layer acts as a geometrical order parameter that vanishes continuously as $\epsilon \to 0$ and obeys a power-law scaling proportional to $\epsilon^{\alpha}$, with exponents $\alpha = 0.496 \pm 0.001$ around $(\pi, 0)$ and $\alpha = 0.496 \pm 0.006$ around $(\pi, \pi)$. The associated response function diverges in the integrable limit, providing clear numerical evidence of critical behavior belonging to the $\alpha \simeq 1/2$ universality class of weakly perturbed Hamiltonian systems.
\end{abstract}


\maketitle

\section{Introduction}
\label{sec:intro}

The transition from regular to chaotic motion is a foundational problem in the theory of nonlinear dynamical systems and Hamiltonian mechanics \cite{zaslavsky1985chaos,hilborn1994chaos,lichtenberg1992,meiss1992synplectic}. In integrable Hamiltonian systems, trajectories are confined to invariant tori that foliate the phase space and prevent large-scale transport along the action coordinate. When a weak nonlinear perturbation is introduced, this ordered foliation is partially broken. According to the Kolmogorov-Arnold-Moser (KAM) theorem, sufficiently irrational invariant tori survive under small perturbations as deformed spanning curves \cite{kolmogorov1954,arnold1963,moser1962}, whereas resonant tori are destroyed through the Poincar\'e-Birkhoff mechanism and replaced by alternating chains of elliptic and hyperbolic periodic points \cite{lichtenberg1992}. Transverse intersections between the stable and unstable manifolds emanating from hyperbolic points create homoclinic and heteroclinic tangles, giving rise to localized stochastic layers \cite{chirikov1979,MacKay1984,mackay1987hamiltonian}. As the perturbation strength increases, these stochastic regions expand and overlap, progressively destroying the remaining invariant spanning curves and leading to a mixed phase space where stability islands, cantori, and sticky regions coexist with chaotic seas \cite{greene1979,mackay1983,contopoulos1971orbits,Zaslavsky2002,Altmann2005,Altmann2006,scher2023escape}.

Area-preserving two-dimensional mappings provide an effective setting for investigating the mechanisms of resonance overlap, anomalous diffusion, and dynamical trapping in conservative systems \cite{Taylor1975,rechester1980calculation,rechester1981fourier,sysm1993,Thanos2014,harsoula2019,borin2023,rolimsales2024escape,rolimsales2024ratchet,borin2024hurst,borin2024caputo,borin2025fractional,borin2026lemon}. Among the physical models described by such discrete mappings, the bouncer model holds particular importance due to its connection with Fermi acceleration \cite{fermi1949,lichtenberg1980,lieberman1998}. Originally formulated to study a classical particle falling in a constant gravitational field and colliding elastically with a periodically oscillating horizontal wall, the model was rigorously analyzed by Pustyl'nikov, who proved the existence of initial conditions leading to unbounded energy growth \cite{pustylnikov1972,pustylnikov1974,kruger1995}. In its simplified static-wall approximation, the bouncer model preserves the essential canonical structure and phase-space topology of the complete system while avoiding the numerical solution of transcendental equations between successive collisions \cite{lichtenberg1980,livorati2008,livorati2012,leonel2015cnsns}. Previous investigations of the conservative and dissipative bouncer mappings have extensively characterized the global chaos threshold, Fermi acceleration, survival probability, finite-time Lyapunov exponents, and the slowing down of energy growth induced by stickiness around KAM islands \cite{livorati2008,livorati2012,borin2024hurst}.

Beyond the traditional geometric analysis of KAM torus destruction and global transport thresholds, recent works have shown that the onset of chaos in conservative systems can be formulated within the framework of continuous phase transitions borrowed from statistical mechanics \cite{frohlich2024,chaikin1995,sethna2021,goldenfeld2018lectures,leonel2015pla,leonel2020,leonel2023dynamical,leonel2026pre,leonel2026epl}. In this approach, the breakdown of the integrable phase-space foliation as the perturbation parameter departs from zero is quantified by a geometrical order parameter defined from the width of the emerging stochastic layer. When this order parameter vanishes continuously at the integrable limit according to a power law with a critical exponent between zero and unity, its derivative with respect to the control parameter defines a response function (or susceptibility) that diverges at the transition point \cite{leonel2026pre,leonel2026epl}. Remarkably, a critical exponent close to $\alpha = 1/2$ has been reported across several distinct Hamiltonian systems, including the Chirikov-Taylor standard map, the symmetric standard map,
 and oval-shaped billiards \cite{leonel2026pre, borin2027dynamical, leonel2026epl, LEONEL2027135476}.

The objective of the present work is to investigate the onset of non-integrability in the conservative bouncer model by characterizing the emergence and scaling of the stochastic layers formed around its fixed points as the perturbation parameter $\epsilon$ is varied. Because the simplified bouncer mapping incorporates a non-smooth absolute-value reflection at zero velocity to prevent the particle from penetrating the wall, the local phase-space geometry near the boundary $V = 0$ differs qualitatively from that near interior fixed points such as $V = \pi$. Consequently, a comparative numerical analysis of the stochastic layers surrounding the fixed points $(\pi, 0)$ and $(\pi, \pi)$ is carried out. By constructing precise numerical boundary curves through an angular binning procedure, the scaling behavior of both stochastic layers is determined over two orders of magnitude of the perturbation parameter near the integrable limit.

This paper is organized as follows. Section~\ref{sec:model} introduces the bouncer model, its area-preserving mapping equations, the stability of its fixed points, and the emergence of microscopic stochastic layers in phase space. Section~\ref{sec:results} details the numerical methodology used to reconstruct the boundary curves and presents the scaling results for the stochastic layers around $(\pi, 0)$ and $(\pi, \pi)$, followed by a comparative discussion of both regimes. Finally, Section~\ref{sec:conclusions} summarizes the main conclusions.

\section{Bouncer Model}
\label{sec:model}

The bouncer model describes the vertical motion of a classical particle of mass $m$ subject to a constant downward gravitational acceleration $g$, undergoing elastic collisions with a heavy horizontal wall oscillating periodically in time with amplitude $\varepsilon$ and angular frequency $w$ according to $y(t) = \varepsilon \cos(w t)$ \cite{lichtenberg1980,livorati2012}. In the complete formulation of the model, tracking the exact collision position requires distinguishing between single and multiple impacts within the collision zone and solving transcendental equations at each iteration \cite{livorati2012}. To simplify the discrete-time evolution while preserving the Hamiltonian transport properties and mixed phase-space structure, the static-wall approximation assumes that the wall remains fixed at $y = 0$ for the flight-time calculation, but transfers momentum to the particle at the instant of impact as though the wall were moving \cite{lichtenberg1980,livorati2008,livorati2012,borin2024hurst}.

Introducing dimensionless variables for the phase angle at the $n$-th collision, $\phi_n = w t_n$, the conjugate action variable (dimensionless particle velocity), $V_n = v_n w / g$, and the dimensionless control parameter governing the nonlinear perturbation, $\epsilon = \varepsilon w^2 / g$, the dynamics of the conservative bouncer model is governed by the two-dimensional mapping $(\phi_n, V_n) \mapsto (\phi_{n+1}, V_{n+1})$ defined as \cite{livorati2012,borin2024hurst}
\begin{align}
\phi_{n+1} &= (\phi_n + 2 V_n) \bmod 2\pi, \label{eq:map_theta} \\
V_{n+1} &= \left| V_n - 2\epsilon \sin(\phi_{n+1}) \right|, \label{eq:map_I}
\end{align}
where $\phi_n \in [0, 2\pi)$ and $V_n \geq 0$. The absolute value in Eq.~\eqref{eq:map_I} enforces the physical constraint that the particle cannot move into the region of negative velocities below the rigid wall; whenever an impact would yield a negative velocity, the particle is immediately reinjected upward with the corresponding positive speed \cite{livorati2012}. Evaluating the Jacobian matrix of Eqs.~\eqref{eq:map_theta} and~\eqref{eq:map_I} yields $\det J = \pm 1$, confirming that the mapping preserves phase-space area and satisfies Liouville's theorem. 

The period-one fixed points $(\phi^*, V^*)$ of the bouncer map are obtained from the condition $(\phi_{n+1}, V_{n+1}) = (\phi_n, V_n) = (\phi^*, V^*)$, which yields \cite{lichtenberg1992,livorati2012}
\begin{equation}
V^* = \pi l, \qquad \phi^* = \arcsin\!\left(\frac{\pi m}{2\epsilon}\right), \label{eq:fixed_points}
\end{equation}
where $l = 0, 1, 2, \dots$ and $m = 0, 1, 2, \dots$. Linear stability analysis shows that these period-one fixed points are stable within parameter windows given by $(2\pi m)^2 < 16\epsilon^2 < 16 + (2\pi m)^2$ \cite{lichtenberg1992,livorati2012}. For the fundamental branch $m = 0$, the fixed points are located at $\phi^* = 0$ and $\phi^* = \pi$ with $V^* = \pi l$. In the regime of weak nonlinearity, $0 < \epsilon < 1$, the points $(\pi, \pi l)$ correspond to elliptic centers surrounded by stability islands, whereas the points $(0, \pi l)$ act as hyperbolic saddles whose intersecting stable and unstable manifolds organize the stochastic layers enveloping the islands. Due to the vertical periodicity $V^* = \pi l$, the fixed point $(\pi, 0)$ lies directly on the reflecting boundary $V = 0$, whereas the fixed point $(\pi, \pi)$ lies in the interior of the action domain at the interface between the regions $V \leq \pi$ and $V \geq \pi$.

\begin{figure*}[t]
\centering
\includegraphics[width=\linewidth]{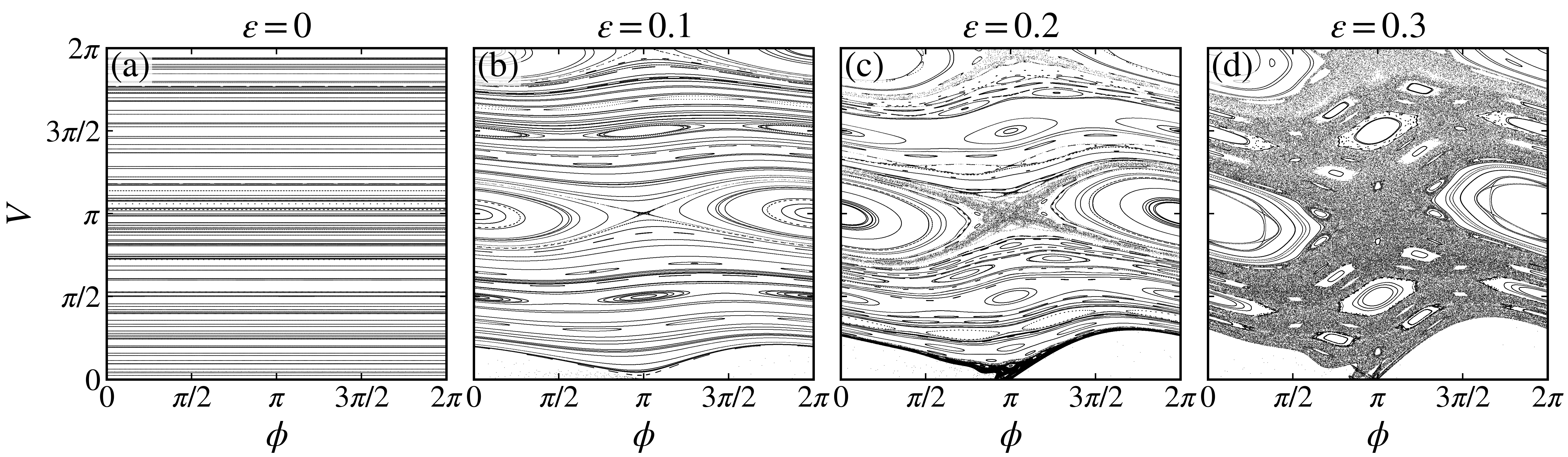}
\caption{Phase-space portraits of the conservative bouncer mapping given by Eqs.~\eqref{eq:map_theta} and~\eqref{eq:map_I} on the domain $(\phi, V) \in [0, 2\pi] \times [0, 2\pi]$ for four values of the perturbation parameter: (a) $\epsilon = 0$, (b) $\epsilon = 0.1$, (c) $\epsilon = 0.2$, and (d) $\epsilon = 0.3$. Each panel is generated by iterating $N_{\mathrm{IC}} = 120$ uniformly distributed random initial conditions for $N = 3 \times 10^3$ steps.}
\label{fig:fig1}
\end{figure*}

Figure~\ref{fig:fig1} illustrates the global evolution of the phase space of Eqs.~\eqref{eq:map_theta} and~\eqref{eq:map_I} over the domain $(\phi, V) \in [0, 2\pi] \times [0, 2\pi]$ for four values of the control parameter: $\epsilon = 0$, $0.1$, $0.2$, and $0.3$. 
In the integrable limit, $\epsilon = 0$ [Fig.~\ref{fig:fig1}(a)], the action variable is an exact constant of motion ($V_{n+1} = V_n$), and the phase space is entirely foliated by horizontal invariant lines corresponding to periodic or quasiperiodic rotations in $\phi$. When the perturbation is turned on at $\epsilon = 0.1$ [Fig.~\ref{fig:fig1}(b)] and $\epsilon = 0.2$ [Fig.~\ref{fig:fig1}(c)], the integrable foliation breaks down: stability islands emerge around the elliptic fixed points at $\phi = \pi$, $V = 0, \pi, 2\pi$, while invariant spanning curves continue to traverse the phase space horizontally, confining chaotic motion to localized layers. Above the critical value $\epsilon_c \approx 0.2429$ \cite{lichtenberg1992,livorati2012}, as shown for $\epsilon = 0.3$ in Fig.~\ref{fig:fig1}(d), the invariant spanning curves are destroyed, and the local stochastic layers merge into a connected chaotic sea that coexists with the surviving stability islands.


\begin{figure*}[t]
\centering
\includegraphics[width=\linewidth]{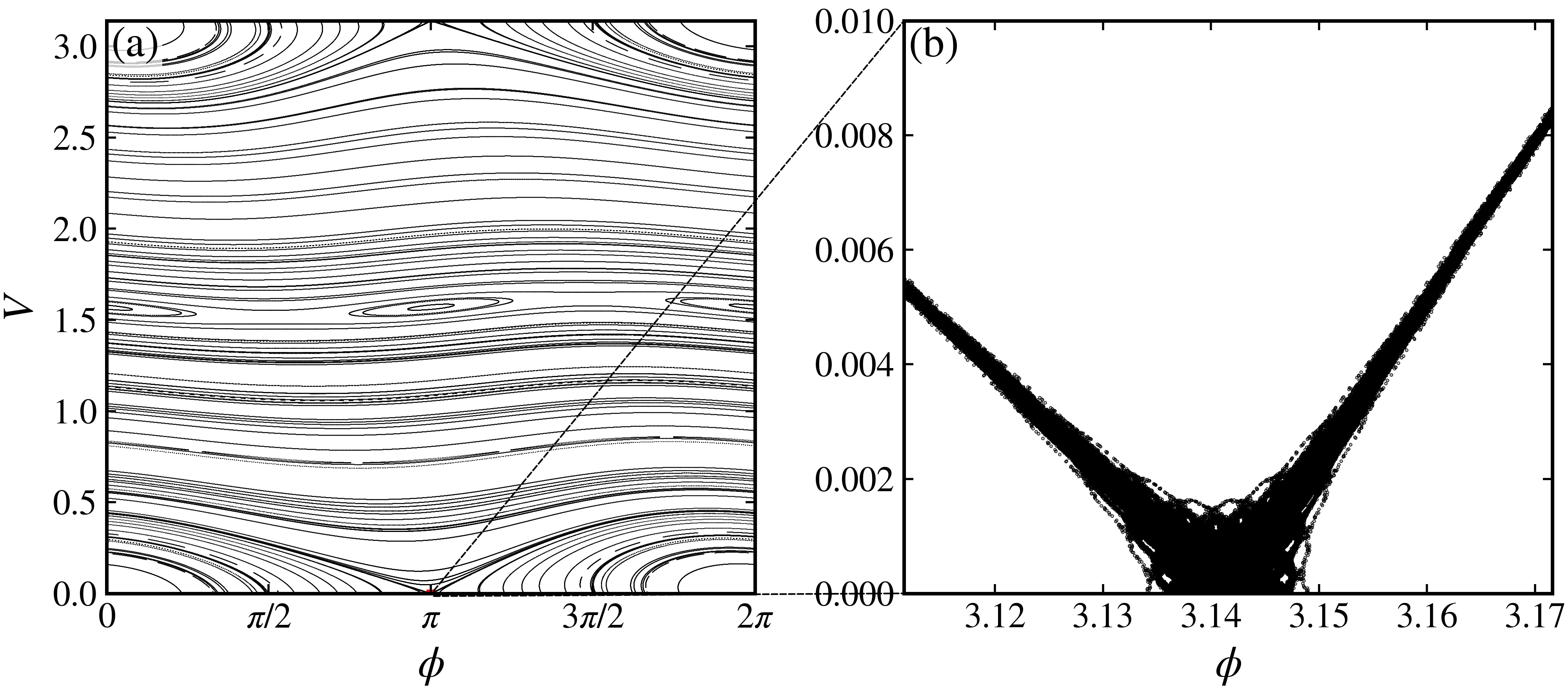}
\caption{
(a)
Phase space of the bouncer model at $\epsilon = 0.05$. 
At this global scale, the phase space appears completely regular. (b) Magnification of the rectangular region around $(\pi, 0)$ over the window $\phi \in [\pi - 0.03, \pi + 0.03]$ and $I \in [0, 0.01]$, revealing a well-developed stochastic layer.
}
\label{fig:fig2}
\end{figure*}

For very small values of the perturbation parameter, a global inspection of the phase space may suggest that the dynamics remains entirely regular and integrable. However, examining the vicinity of the fixed points at higher resolution reveals that localized chaotic behavior is already present as soon as $\epsilon > 0$. This phenomenon is demonstrated in Fig.~\ref{fig:fig2} for $\epsilon = 0.05$.
Figure~\ref{fig:fig2}(a) shows the phase space  in the $(\phi, V)$ plane. At this scale, the trajectories appear to follow smooth, regular curves with no visible sign of chaos. 
Figure~\ref{fig:fig2}(b) displays a magnified view of 
a
region surrounding $(\pi, 0)$ within the window 
$\phi \in [\pi - 0.03, \pi + 0.03]$ and $V \in [0, 0.01]$
, where $100$ trajectories initialized in the microscopic neighborhood $\delta_{\mathrm{micro}} = 10^{-12}$ are evolved for $N = 5 \times 10^3$ iterations. The magnification clearly exposes a band of scattered points occupying a finite area around $(\pi, 0)$, confirming the existence of a narrow stochastic layer close to the integrable limit.

\section{Results}
\label{sec:results}

To quantitatively characterize the transition from integrability to chaos in the bouncer model, the numerical procedures used to reconstruct the boundaries of the stochastic layers and determine their scaling with the perturbation parameter $\epsilon$ are presented below. 
Two distinct analyses are performed, focusing on the stochastic layers surrounding the fixed points $(\pi, 0)$ and $(\pi, \pi)$.

\subsection{Stochastic layer around the fixed point $(\pi, 0)$}
\label{sec:pi0}

Around the fixed point $(\pi, 0)$, the non-negativity of the action variable in Eq.~\eqref{eq:map_I} confines the stochastic layer entirely to the upper half-plane $V \geq 0$. As a chaotic trajectory wanders through this layer, its vertical excursion is bounded from below by the regular island core surrounding $(\pi, 0)$ (and the line $V = 0$ near the hyperbolic points) and from above by the first surviving invariant curve. Consequently, the geometry of the stochastic layer can be described by two boundary curves: a lower boundary $\gamma_{\min}(\phi)$ and an upper boundary $\gamma_{\max}(\phi)$.

To construct these curves numerically for a given perturbation parameter $\epsilon$, a single chaotic trajectory is initialized in the immediate vicinity of the fixed point at
$
(\phi_0, V_0) = (\pi, 10^{-6}) \label{eq:ic_pi0}
$
and iterated $N$ times according to Eqs.~\eqref{eq:map_theta} and~\eqref{eq:map_I}
.
The angular interval $[0, 2\pi]$ is partitioned into $N_{\mathrm{bins}}$ uniform bins of width $\Delta\phi = 2\pi / N_{\mathrm{bins}}$, defined by the bin edges $\phi_{\mathrm{edge},k} = k \Delta\phi$ for $k = 0, 1, \dots, N_{\mathrm{bins}}$ and the corresponding bin centers
\begin{equation}
\phi_j = \left(j - \frac{1}{2}\right)\Delta\phi, \qquad j = 1, 2, \dots, N_{\mathrm{bins}}. \label{eq:bin_centers}
\end{equation}
Each iterate $(\phi_i, V_i)$ of the trajectory is assigned to its respective angular bin $j$. Within each bin $j$, the minimum and maximum action values reached by the orbit are extracted independently as
\begin{align}
\gamma_{\min,j}(\epsilon) &= \min_{\phi_i \in \text{bin } j} V_i, \label{eq:gamma_min_pi0} \\
\gamma_{\max,j}(\epsilon) &= \max_{\phi_i \in \text{bin } j} V_i. \label{eq:gamma_max_pi0}
\end{align}
Associating the discrete extrema $\gamma_{\min,j}(\epsilon)$ and $\gamma_{\max,j}(\epsilon)$ with the bin center $\phi_j$ defines the numerical boundary curves $\gamma_{\min}(\phi)$ and $\gamma_{\max}(\phi)$. 

\begin{figure*}[t]
\centering
\includegraphics[width=\linewidth]{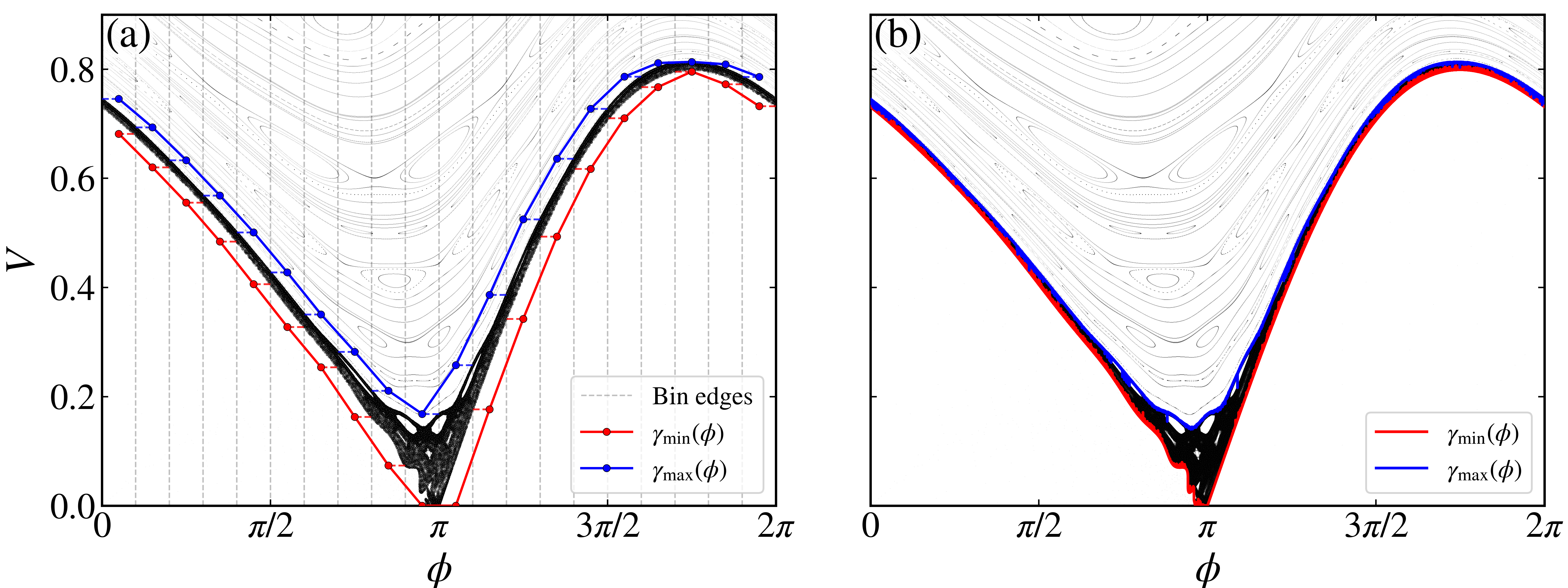}
\caption{Numerical construction of the lower boundary $\gamma_{\min}(\phi)$ (red) and upper boundary $\gamma_{\max}(\phi)$ (blue) of the stochastic layer around $(\pi, 0)$ for $\epsilon = 0.15$. 
(a) Low-resolution construction using $N_{\mathrm{bins}} = 20$ angular bins and $N = 10^5$ iterations; vertical dashed lines mark the bin edges and horizontal dashed segments connect the exact extremum points to the bin centers. (b) High-resolution construction using $N_{\mathrm{bins}} = 10^4$ angular bins and $N = 10^7$ iterations, accurately delineating the inner and outer envelopes of the stochastic layer.}
\label{fig:fig3}
\end{figure*}

Figure~\ref{fig:fig3} illustrates the boundary extraction procedure at $\epsilon = 0.15$ and demonstrates the role of the angular bin resolution. 
In Fig.~\ref{fig:fig3}(a), a coarse partition of $N_{\mathrm{bins}} = 20$ angular bins and $N = 10^5$ iterations is used to make the mechanics of the binning algorithm explicitly visible: the vertical gray dashed lines indicate the bin boundaries, and the horizontal red and blue dashed segments show how the minimum and maximum action values found anywhere within bin $j$ are projected horizontally onto the bin center $\phi_j$. Because each bin spans a wide angular interval ($\Delta\phi = \pi/10$) over which the true boundary curves vary appreciably, the coarse discretization in Fig.~\ref{fig:fig3}(a) underestimates the local inner boundary and overestimates the local outer boundary whenever the envelope has a steep slope. In contrast, Fig.~\ref{fig:fig3}(b) shows the same construction using a high resolution of $N_{\mathrm{bins}} = 10^4$ angular bins and $N = 10^7$ iterations. At this fine angular resolution ($\Delta\phi = 2\pi \times 10^{-4}$), the discrete bin extrema converge to smooth curves $\gamma_{\min}(\phi)$ and $\gamma_{\max}(\phi)$ that tightly trace the complex geometrical contours of both the central stability island and the outer invariant spanning curve bounding the stochastic layer.

To quantify how the thickness of the stochastic layer scales with the perturbation parameter $\epsilon$, the local width in the $j$-th angular bin is defined as
\begin{equation}
\Delta\gamma_j(\epsilon) = \gamma_{\max,j}(\epsilon) - \gamma_{\min,j}(\epsilon). \label{eq:delta_gamma_j}
\end{equation}
The characteristic width of the stochastic layer is then obtained by averaging $\Delta\gamma_j(\epsilon)$ over the occupied angular bins:
\begin{equation}
\langle\Delta\gamma\rangle = \frac{1}{N_{\mathrm{bins}}} \sum_{j=1}^{N_{\mathrm{bins}}} \Delta\gamma_j(\epsilon), \label{eq:mean_delta_gamma}
\end{equation}
where the normalization is evaluated over all valid bins containing orbit points (which equals $N_{\mathrm{bins}} = 10^4$ when every angular bin is visited by the trajectory).

\begin{figure}[t]
\centering
\includegraphics[width=0.65\linewidth]{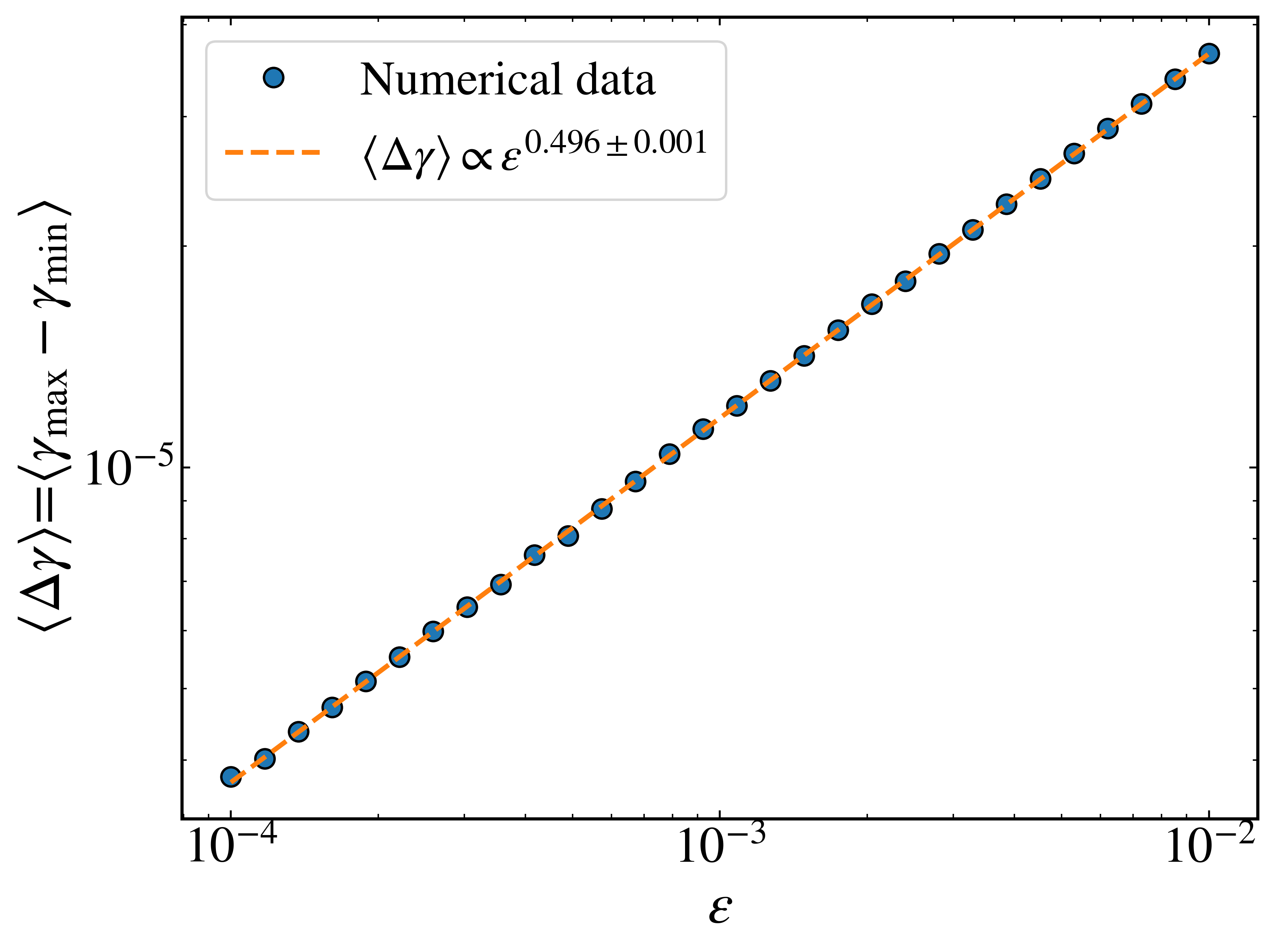}
\caption{Log-log plot of the mean stochastic-layer width $\langle\Delta\gamma\rangle = \langle\gamma_{\max} - \gamma_{\min}\rangle$ around the fixed point $(\pi, 0)$ as a function of the perturbation parameter $\epsilon$. For each $\epsilon$ value,
a single trajectory initialized at $(\phi_0, V_0) = (\pi, 10^{-6})$ is iterated for $N = 10^6$ steps across $N_{\mathrm{bins}} = 10^4$ angular bins. The dashed line represents the linear least-squares power-law fit yielding $\langle\Delta\gamma\rangle \propto \epsilon^{0.496 \pm 0.001}$.}
\label{fig:fig4}
\end{figure}

Figure~\ref{fig:fig4} presents the behavior of $\langle\Delta\gamma\rangle$ as a function of $\epsilon$ near the integrable limit. 
For each value of $\epsilon$, a single trajectory initialized at $(\phi_0, V_0) = (\pi, 10^{-6})$ is iterated for $N = 10^6$ steps and partitioned into $N_{\mathrm{bins}} = 10^4$ uniform angular bins on $[0, 2\pi]$. As shown in the log-log representation of Fig.~\ref{fig:fig4}, the numerical data follow a robust power-law scaling over the entire two-decade interval:
\begin{equation}
\langle\Delta\gamma\rangle \propto \epsilon^{\alpha}. \label{eq:powerlaw_pi0}
\end{equation}
Performing an ordinary linear least-squares regression of $\log_{10}\langle\Delta\gamma\rangle$ against $\log_{10}\epsilon$ over $\epsilon \in [10^{-4}, 10^{-2}]$, with the parameter uncertainty determined from the standard error
of the covariance matrix of the fit, yields the critical exponent
\begin{equation}
\alpha = 0.496 \pm 0.001. \label{eq:alpha_pi0}
\end{equation}
Because $\langle\Delta\gamma\rangle$ vanishes continuously as $\epsilon \to 0$ and is finite for $\epsilon > 0$, it serves as a natural order parameter for the transition from integrability to chaos. The associated response function (susceptibility), defined as the derivative of the order parameter with respect to the control parameter, scales as
\begin{equation}
\chi = \frac{d\langle\Delta\gamma\rangle}{d\epsilon} \propto \epsilon^{\alpha - 1}. \label{eq:chi_pi0}
\end{equation}
Since $0 < \alpha < 1$, the response function diverges as $\chi \to \infty$ in the integrable limit $\epsilon \to 0$, confirming that the emergence of the stochastic layer around $(\pi, 0)$ exhibits the scaling signatures of a continuous dynamical phase transition.

\subsection{Stochastic layer around the fixed point $(\pi, \pi)$}
\label{sec:pipi}

A different numerical procedure is required to characterize the stochastic layer surrounding the interior fixed point $(\pi, \pi)$. Unlike the fixed point $(\pi, 0)$, which sits on the reflecting boundary $V = 0$ so that its stochastic layer is restricted to $V \geq 0$, the fixed point $(\pi, \pi)$ lies at the line $V = \pi$ separating two symmetric regions of the phase space. Consequently, a chaotic trajectory initialized near $(\pi, \pi)$ explores a stochastic layer that branches into two distinct domains: an upper region located at $V \geq \pi$ and a lower region located at $V \leq \pi$, both wrapping around the central stability island centered at $(\pi, \pi)$. If only the global minimum and global maximum of $V$ were extracted in each angular bin, their difference near $\phi = \pi$ would measure the vertical diameter of the regular stability island rather than the thickness of the chaotic layer itself. To isolate the stochastic layer from the regular island core, four boundary curves must be constructed to track the inner and outer envelopes of the upper ($V \geq \pi$) and lower ($V \leq \pi$) branches independently.

In accordance with the numerical implementation, all four boundaries are reconstructed from a single chaotic orbit initialized at
$
(\phi_0, V_0) = (\pi, \pi + 10^{-9}) \label{eq:ic_pipi}
$
and iterated for $N$ steps via Eqs.~\eqref{eq:map_theta} and~\eqref{eq:map_I}. The angular domain $[0, 2\pi]$ is divided into $N_{\mathrm{bins}}$ equal bins centered at $\phi_j$ [Eq.~\eqref{eq:bin_centers}]. Within each angular bin $j$, the trajectory points $(\phi_i, V_i)$ are separated into two subsets according to whether $V_i \geq \pi$ or $V_i \leq \pi$. The four boundary curves are then defined by computing the minimum and maximum action values independently in each subset for every bin $j$:
\begin{align}
\gamma_{\min,j}^{+\pi}(\phi) &= \min_{\phi_i \in \text{bin } j} \{ V_i \geq \pi \}, \label{eq:g_min_plus} \\
\gamma_{\max,j}^{+\pi}(\phi) &= \max_{\phi_i \in \text{bin } j} \{ V_i \geq \pi \}, \label{eq:g_max_plus} \\
\gamma_{\min,j}^{-\pi}(\phi) &= \min_{\phi_i \in \text{bin } j} \{ V_i \leq \pi \}, \label{eq:g_min_minus} \\
\gamma_{\max,j}^{-\pi}(\phi) &= \max_{\phi_i \in \text{bin } j} \{ V_i \leq \pi \}. \label{eq:g_max_minus}
\end{align}
Here, $\gamma_{\min}^{+\pi}(\phi)$ and $\gamma_{\max}^{+\pi}(\phi)$ represent the lower and upper boundaries of the stochastic branch above $V = \pi$, while $\gamma_{\min}^{-\pi}(\phi)$ and $\gamma_{\max}^{-\pi}(\phi)$ represent the lower and upper boundaries of the stochastic branch below $V = \pi$. 

\begin{figure*}[t]
\centering
\includegraphics[width=\linewidth]{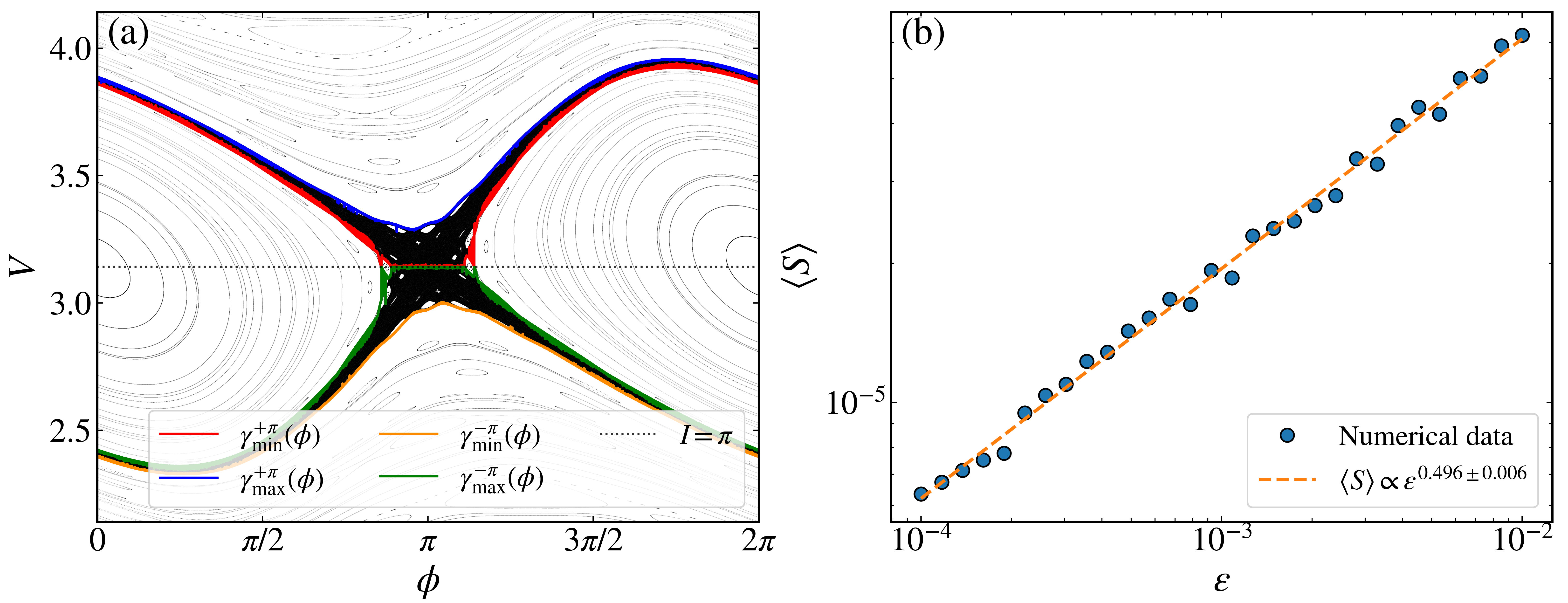}
\caption{Characterization and scaling of the stochastic layer around the fixed point $(\pi, \pi)$. (a) Phase-space portrait on $(\phi, V) \in [0, 2\pi] \times [\pi - 1, \pi + 1]$ for $\epsilon = 0.15$, showing the background trajectories 
(light gray), the single boundary trajectory initialized at $(\phi_0, V_0) = (\pi, \pi + 10^{-9})$ and iterated for $N = 10^6$ steps (dark gray), the dividing line $V = \pi$ (black dotted), and the four boundary curves $\gamma_{\min}^{+\pi}(\phi)$ (red), $\gamma_{\max}^{+\pi}(\phi)$ (blue), $\gamma_{\min}^{-\pi}(\phi)$ (orange), and $\gamma_{\max}^{-\pi}(\phi)$ (green) computed with $N_{\mathrm{bins}} = 10^4$ angular bins. (b) Log-log plot of the mean combined width $\langle S \rangle$ as a function of $\epsilon$ 
, together with the power-law fit yielding $\langle S \rangle \propto \epsilon^{0.496 \pm 0.006}$.}
\label{fig:fig5}
\end{figure*}

Figure~\ref{fig:fig5}(a) displays the phase space around $(\pi, \pi)$ and the four reconstructed boundary curves for $\epsilon = 0.15$. 
The four boundary curves are extracted from the single orbit starting at $(\phi_0, V_0) = (\pi, \pi + 10^{-9})$ using $N = 10^6$ iterations and $N_{\mathrm{bins}} = 10^4$ angular bins. As seen in Fig.~\ref{fig:fig5}(a), the curves $\gamma_{\max}^{+\pi}(\phi)$ and $\gamma_{\min}^{-\pi}(\phi)$ trace the outermost invariant spanning curves above and below $V = \pi$, respectively, while $\gamma_{\min}^{+\pi}(\phi)$ and $\gamma_{\max}^{-\pi}(\phi)$ outline the upper and lower borders of the stability islands embedded along the line $V = \pi$.

Because the stochastic layer around $(\pi, \pi)$ is composed of both the upper and lower branches, the local thickness of the chaotic region in the $j$-th angular bin is quantified by the sum of the widths of the two branches:
\begin{equation}
S_j(\epsilon) = \left| \gamma_{\max,j}^{+\pi}(\epsilon) - \gamma_{\min,j}^{+\pi}(\epsilon) \right| + \left| \gamma_{\max,j}^{-\pi}(\epsilon) - \gamma_{\min,j}^{-\pi}(\epsilon) \right|. \label{eq:S_j}
\end{equation}
The mean combined width $\langle S \rangle$ is then evaluated by averaging $S_j(\epsilon)$ strictly over the subset of valid angular bins $\mathcal{V}$ in which all four boundary quantities $\gamma_{\min,j}^{+\pi}$, $\gamma_{\max,j}^{+\pi}$, $\gamma_{\min,j}^{-\pi}$, and $\gamma_{\max,j}^{-\pi}$ are simultaneously defined and finite:
\begin{equation}
\langle S \rangle = \frac{1}{N_{\mathrm{bins}}} \sum_{j=1}^{N_{\mathrm{bins}}}  S_j(\epsilon), \label{eq:mean_S}
\end{equation}

Figure~\ref{fig:fig5}(b) shows the dependence of $\langle S \rangle$ on the perturbation parameter $\epsilon$ in log-log scale. 
Over the entire scaling interval, the mean combined width obeys the power law
$
\langle S \rangle \propto \epsilon^{\alpha}, \label{eq:powerlaw_pipi}
$
with 
$
\alpha = 0.496 \pm 0.006. \label{eq:alpha_pipi}
$
As in the previous case, the order parameter $\langle S \rangle$ approaches zero continuously as $\epsilon \to 0$, while its associated response function $\chi = d\langle S \rangle / d\epsilon \propto \epsilon^{\alpha - 1}$ diverges in the integrable limit.

\subsection{Comparison of the two scaling analyses}
\label{sec:comparison}

Comparing the results obtained for the fixed points $(\pi, 0)$ and $(\pi, \pi)$ reveals a notable agreement in their critical scaling behavior despite the distinct geometrical procedures required to measure each stochastic layer. Around $(\pi, 0)$, the presence of the rigid wall at $V = 0$ folds negative post-collision velocities back into the positive domain through the modulus operation in Eq.~\eqref{eq:map_I}. As a result, the stochastic layer is one-sided and completely enclosed between a single lower boundary $\gamma_{\min}(\phi)$ and a single upper boundary $\gamma_{\max}(\phi)$, allowing its transverse extent to be quantified directly by $\langle\Delta\gamma\rangle$. In contrast, around the interior fixed point $(\pi, \pi)$, the stochastic layer develops on both sides of the line $V = \pi$ and surrounds the central stability island. Capturing the true chaotic thickness without including the regular island interior requires partitioning the phase space at $V = \pi$, extracting four independent boundary curves, and defining the composite quantifier $\langle S \rangle$ over bins populated in both regions.

Despite these differences in local phase-space topology and boundary reconstruction, both analyses yield identical power-law exponents within numerical uncertainty: $\alpha = 0.496 \pm 0.001$ for $(\pi, 0)$ and $\alpha = 0.496 \pm 0.006$ for $(\pi, \pi)$. The slightly larger uncertainty of $\pm 0.006$ in the $(\pi, \pi)$ analysis arises naturally from the stricter condition that all four boundaries must be simultaneously populated within a bin and from the summation of fluctuations across two separate branches. Crucially, both values are in excellent agreement with the theoretical and numerical value $\alpha \simeq 1/2$ previously established for the Chirikov-Taylor standard map ($\alpha = 0.5004 \pm 0.0002$) \cite{LEONEL2027135476}, the symmetric standard map ($\alpha = 0.513 \pm 0.004$) \cite{borin2027dynamical}, 
and oval billiards \cite{leonel2026pre,leonel2026epl}. This consistency demonstrates that the non-smooth reflection at $V = 0$ does not alter the critical exponent governing the growth of the stochastic layer, confirming that the onset of chaotic transport in the bouncer model belongs to the same universality class of continuous dynamical phase transitions.

\section{Conclusions}
\label{sec:conclusions}

The transition from integrability to chaos in the conservative bouncer model has been investigated by analyzing the emergence and scaling of the stochastic layers formed around the period-one fixed points $(\pi, 0)$ and $(\pi, \pi)$. In the integrable limit, $\epsilon = 0$, the action is conserved and the phase space is foliated by horizontal invariant lines. For any non-zero perturbation, $\epsilon > 0$, this foliation is broken and thin stochastic layers develop in the vicinity of the fixed points, even when global phase-space portraits appear regular at macroscopic scales.

By dividing the angular coordinate into $N_{\mathrm{bins}} = 10^4$ bins and recording the action extrema of a single chaotic trajectory, the boundary curves of the stochastic layers were reconstructed with high precision. For the fixed point $(\pi, 0)$, the layer lies entirely in $V \geq 0$ and is characterized by two boundary curves, $\gamma_{\min}(\phi)$ and $\gamma_{\max}(\phi)$, whose mean separation scales over $\epsilon \in [10^{-4}, 10^{-2}]$ as $\langle\Delta\gamma\rangle \propto \epsilon^{\alpha}$ with $\alpha = 0.496 \pm 0.001$. For the interior fixed point $(\pi, \pi)$, where the stochastic layer extends into two branches separated by $V = \pi$, a four-boundary procedure was introduced to evaluate the combined width $\langle S \rangle$, yielding $\langle S \rangle \propto \epsilon^{\alpha}$ with $\alpha = 0.496 \pm 0.006$.

In both regions of the phase space, the characteristic width of the stochastic layer acts as a geometrical order parameter that vanishes continuously at $\epsilon = 0$, while the corresponding response function $\chi \propto \epsilon^{\alpha - 1}$ diverges as $\epsilon \to 0$. The coincidence of the exponents $\alpha \simeq 1/2$ across both fixed points, and their agreement with those observed in standard mappings and billiard systems, provides strong evidence that the onset of non-integrability in weakly perturbed conservative systems is governed by a universal square-root scaling law.

\section*{Acknowledgments}

This study was financed, in part, by the S\~ao Paulo Research Foundation (FAPESP), Brazil, Process Numbers 2026/08008-0 and 2023/08698-9. E.D.L. acknowledges support from Brazilian agencies CNPq (No. 304398/2023-3) and FAPESP (No. 2025/14544-0, No. 2025/27957-0).

\bibliography{references}

\end{document}